# Improving approximate-optimized effective potentials by imposing exact conditions: Theory and applications to electronic statics and dynamics

Yair Kurzweil and Martin Head-Gordon

*Department of Chemistry, University of California at Berkeley, and,*

*Chemical Sciences Division, Lawrence Berkeley National Laboratory,*

*Berkeley, CA 94720, USA*

We develop a method that can constrain any local exchange-correlation potential to preserve basic exact conditions. Using the method of Lagrange multipliers, we calculate for each set of given Kohn-Sham orbitals, a constraint-preserving potential which is closest to the given exchange-correlation potential. The method is applicable to both the time-dependent (TD) and independent cases. The exact conditions that are enforced for the time-independent case are: Galilean covariance, zero net force and torque, and Levy-Perdew virial theorem. For the time-dependent case we enforce translational covariance, zero net force, the Levy-Perdew virial theorem and energy balance. We test our method on the exchange (only) Krieger-Li-Iafrate (xKLI) approximate-optimized effective potential (AOEP), for both cases. For the time-independent case, we calculated the ground state properties of some hydrogen chains and small sodium clusters, for some constrained xKLI potentials and Hartree-Fock (HF) exchange. The results (total energy, Kohn-Sham eigenvalues, polarizability and hyperpolarizability ) indicate that enforcing the exact conditions is not important for these cases. On the other hand, in the time-dependent case, constraining both energy balance and zero net force yields improved results relative to TDHF calculations. We explored the electron dynamics in small sodium clusters driven by CW laser pulses. For each laser pulse we compared calculations from TD constrained xKLI, TD partially constrained xKLI and TDHF. We found that electron dynamics like electron ionization and moment of inertia dynamics for the constrained xKLI are most similar to the TDHF results. Also, energy conservation is better by at least an order of magnitude with respect to the unconstrained xKLI. We also discuss the problems that arise in satisfying constraints in the TD case with a non CW driving force.

I. INTRODUCTION

Density functional theory (DFT)[1-3] and time-dependent density functional theory (TDDFT)[4] are powerful and yet practical theories that are able to predict both ground state and dynamical properties of many-electron molecules. The method of mapping interacting systems onto noninteracting systems results in a huge reduction of computational expense, due to the change from a many-body Hamiltonian to an effective single particles Hamiltonian in the Kohn-Sham (KS) method.[2,5] The main price, however, is the need to approximate the unknown exchange-correlation (XC) functional.[2-4] The simplest approximation is the local density approximation (LDA),[3,6] where the XC functional is a *function* of the local density. While LDA and its improved versions (involving spatial derivatives of the density) are very successful in many applications,[3] they suffer from some fundamental problems,[6] e.g. incorrect asymptotic behavior, missing the derivative discontinuity of the energy near integer number of particles, self-interaction, and a lack of systematic improvement beyond the homogeneous electron gas.

A potential advance toward more realistic functionals is to approximate the XC energy as a functional of the KS orbitals, rather than the density.[7] Obtaining the corresponding local XC potential in this case is done within the framework of the optimized effective potential (OEP) method.[8,9] The basic equation for calculating the OEP involves an integral equation, which has until very recently been found to be numerically highly expensive. Therefore, some practical approximated OEP (AOEP) methods have been developed[7,10-13] (On the other hand, there is currently a new method that makes the *time-independent* OEP evaluation stable and feasible[14]). AOEP's have also been extended to the TD case,[15] where TD-OEP[7,16,17] is still considered a very demanding calculation. Recent experimental developments in experimental ultrafast spectroscopy[18-21] (are opening new frontiers where simulating time-dependent electronic dynamics is of growing importance. It is therefore of interest to assess and further improve AOEP's for this and other purposes.

One significant drawback of these (TD-) AOEPs potentials is the fact that they are not a complete functional derivative of any parent functional.[22] Therefore, the Hellman-Feynman theorem does not hold. As a further result, some known exact conditions are not satisfied. The sometimes severe consequences of violating these exact conditions, for TD cases, are discussed and demonstrated in Refs.[22,23] It is the aim of this work to correct any given AOEP to preserve exact conditions for both TD and time-independent cases. Our method, following the work of Kurzweil and Baer,[23,24] uses the method of Lagrange multipliers in order to enforce desired exact conditions on a constraint-preserving new XC potential, optimally close to the given AOEP. The violation of known constraints is a symptom of the underlying failure of AOEP's to be a functional derivative, rather than the fundamental problem itself. Therefore the main question that this paper tries to address is whether correcting an AOEP to eliminate constraint violations is also sufficient to significantly improve calculated observables. To explore this question, we test our method for the case of the exchange only KLI (xKLI)[10] potential for both DFT and TDDFT. We show the changes associated with constraint satisfaction on the static polarizabilities of hydrogen chains, and on electron dynamics in small sodium clusters.

The paper is organized as follows. In Sec. II we first briefly recall the principles of DFT and some exact conditions which the XC potential must satisfy. We describe our method for constraining approximate potentials to satisfy these exact conditions, and show the effect on polarizabili-ties of hydrogen chains. In Sec. III we describe an extended version of the method for TDDFT, and discuss the numerical implementations and its difficulties. Applications to electron dynamics in $Na_5$ driven by CW laser pulses are then presented, followed by some discussion and conclusions.

## II. CONSTRUCTION OF TIME-INDEPENDENT CONSTRAINT-PRESERVING XC POTENTIALS

### A. SOME FUNDAMENTAL EXACT CONDITIONS IN DENSITY FUNCTIONAL THEORY

Assume a system of $N_e$ electrons bound in a static external potential $v_{ext}$. According to the Hohenberg-Kohn[1] theorem, their total energy functional can be written as

$$E[n] = T_S[n] + V_{ext}[n] + E_H[n] + E_{XC}[n], \quad (1)$$

where

$$\begin{aligned}
T_S &= \sum_{i,\sigma} \left\langle \varphi_i^\sigma \left| -\tfrac{1}{2}\nabla^2 \right| \varphi_i^\sigma \right\rangle, \\
V_{ext}[n] &= \int dr\, v_{ext}(\mathbf{r}) n(\mathbf{r}), \\
E_H[n] &= \iint dr\, dr'\, n(\mathbf{r}') n(\mathbf{r})/|\mathbf{r}-\mathbf{r}'|, \\
E_{XC}[n] &\equiv \left\langle \Psi[n] \left| \hat{T}+\hat{W} \right| \Psi[n] \right\rangle - E_H[n],
\end{aligned} \quad (2)$$

$\hat{T}$ is the many body kinetic energy operator, $\hat{W}$ is the 2-body Coulomb interaction operator and $\Psi$ is the exact wavefunction of the interacting electrons. The minima of $E[n]$ can be obtained from the Kohn-Sham (KS) equations:[2]

$$\left[ -\tfrac{1}{2}\nabla^2 + v_{ext}(\mathbf{r}) + v_H(\mathbf{r}) + v_{XC}^\sigma(\mathbf{r}) \right] \varphi_i^\sigma(\mathbf{r}) = \varepsilon_i^\sigma \varphi_i^\sigma(\mathbf{r}), \quad (3)$$

where $v_H(\mathbf{r}) = \int dr'\, n(\mathbf{r}')/|\mathbf{r}-\mathbf{r}'|$ and $v_{XC}(\mathbf{r}) \equiv \delta E_{XC}[n]/\delta n(\mathbf{r})$.

From the $1/|\mathbf{R}-\mathbf{R}'|$ nature of the Coulomb interaction, it can be shown that the XC energy functional is invariant under translations or rotations of the coordinates, which is called Galilean invariance (GI).[25] Mathematically, GI of the XC energy functional, $E_{XC}[n] \equiv E_X[n] + E_C[n]$, means that given two coordinate systems, $\mathbf{r}$ and $\mathbf{r}'$, the energy functional is invariant under the transformation

$$E_{X,C}[n';\mathbf{r}'] = E_{X,C}[n;\mathbf{r}], \quad (4)$$

where $\mathbf{r}' = \mathbf{r} + \mathbf{x}$ for translations or $\mathbf{r}' = M\mathbf{r}$ for rotations. GI implies immediately additional four properties of the derived XC *potential*, $v_{X,C}[n](\mathbf{r}) \equiv \delta E_{X,C}[n]/\delta n(\mathbf{r})$. The first is translational covariance (TC):[25]

$$v_{X,C}[n'](\mathbf{r}) = v_{X,C}[n](\mathbf{r} + \mathbf{x}), \quad (5)$$

and the second is rotational covariance (RC):[25]

$$v_{X,C}[n'](\mathbf{r}) = v_{X,C}[n](M\mathbf{r}), \quad (6)$$

We describe a functional which satisfies both TC and RC as a Galilean *covariant* (GC) functional. Since (4) is valid for any translation or rotation, it is easy to show, together with the Perdew-Levy virial theorems[26] (see below), that for infinitesimal translation or rotation one can obtain the XC zero-force (ZF) and XC zero-torque (ZT) conditions:[25]

$$\begin{aligned}\mathbf{f}_{X,C}[n](\mathbf{r}) &\equiv -\sum_\sigma \int d\mathbf{r}\, n_\sigma(\mathbf{r}) \nabla v^\sigma_{X,C}(\mathbf{r}) = \mathbf{0}, \\ \mathbf{t}_{X,C}[n](\mathbf{r}) &\equiv -\sum_\sigma \int d\mathbf{r}\, n_\sigma(\mathbf{r}) \mathbf{r} \times \nabla v^\sigma_{X,C}(\mathbf{r}) = \mathbf{0},\end{aligned} \quad (7)$$

where $\sigma$ is the spin component for spin polarized systems, $n = \sum_\sigma n_\sigma$, and $v^\sigma_{X,C} \equiv \delta E[n]/\delta n_\sigma$. Other known properties of the exact XC functional are the Perdew-Levy virial theorems (VT) (for any density):[26]

$$E_X[n] + \sum_\sigma \int d\mathbf{r}\, n(\mathbf{r}) \mathbf{r} \cdot \nabla v^\sigma_{X,C}(\mathbf{r}) = 0, \quad (8)$$

$$\begin{aligned}&E_C[n] + \sum_\sigma \int d\mathbf{r}\, n_\sigma(\mathbf{r}) \mathbf{r} \cdot \nabla v^\sigma_{X,C}(\mathbf{r}) + \\ &T_S[n] - T[n] = 0\end{aligned} \quad (9)$$

where $T_S[n]$ and $T[n]$ are the noninteracting and interacting kinetic energy functionals, respectively.

## B. IMPOSING EXACT CONDITIONS ON A GIVEN XC POTENTIAL

In general, an approximate optimized effective potential (AOEP) will not be the complete functional derivative of any density/orbital functional, and therefore may not satisfy the ZF, ZT and the VT conditions. This situation is common to AOEPs where the XC energy functional is orbital-dependent.[7, 22] On the other hand, they may certainly be GC, thanks to their orbital-dependency. However, this property alone does not guarantee the other conditions. Let us be *given* such an AOEP, $w^\sigma_{XC}[\{\varphi_i\}](\mathbf{r})$, which is an approximate functional derivative of an orbital dependent functional $E_{X,C}[\{\varphi^\sigma_i\}]$:

$$\delta E_{X,C} / \delta n_\sigma(\mathbf{r}) \approx w_{X,C}^\sigma [\{\varphi_i^\sigma\}](\mathbf{r}). \qquad (10)$$

We would like to impose as many exact conditions as possible in order to correct $w_{X,C}^\sigma(\mathbf{r})$. We wish to construct constrained potentials, denoted as $v_{X,C}^\sigma(\mathbf{r})$, which *exactly* satisfy conditions (5)-(9). In order to keep $v_{X,C}^\sigma(\mathbf{r})$ as close as possible to their original $w_{X,C}^\sigma(\mathbf{r})$, we formulate a constrained variational problem with the Lagrangian (for either the exchange, 'X' subscript, or the correlation, 'C' subscript):

$$L_{X,C} = \tfrac{1}{2} \sum_\sigma \int dr \left| v_{X,C}^\sigma(\mathbf{r}) - w_{X,C}^\sigma(\mathbf{r}) \right|^2 + \vec{A}^{X,C} \cdot \vec{C}^{X,C}, \qquad (11)$$

where $\vec{C}^X$ and $\vec{C}^C$ are vectors that contain the constraints for the exchange and correlation respectively, and $\vec{A}^{X,C}$ are vectors that contain their corresponding Lagrange multipliers. The potentials $v_{X,C}^\sigma(\mathbf{r})$ are obtained from the stationary point of the Lagrangians in Eq. (11), for *given* fixed orbitals and potentials, $w_{X,C}^\sigma$, where the variation is done with respect to $v_{X,C}^\sigma(\mathbf{r})$ only, i.e.

$$\frac{\delta L_{X,C}}{\delta v_{X,C}^\sigma(\mathbf{r})} = v_{X,C}^\sigma(\mathbf{r}) - w_{X,C}^\sigma(\mathbf{r}) + \vec{A}^{X,C} \cdot \frac{\delta \vec{C}^{X,C}}{\delta v_{X,C}^\sigma(\mathbf{r})} = 0, \qquad (12)$$

(For simplicity we assume the same multipliers $\{A_i^{X,C}\}$ for each spin component $\sigma$). Additional variations with respect to the $\{A_i^{X,C}\}$ yield the additional equations:

$$\frac{\partial L_{X,C}}{\partial A_i^{X,C}} = C_i^{X,C} = 0. \qquad (13)$$

(Note that this method is an extension of the method of Kurzweil and Baer, from Ref. [23]).

The construction of the constraints $\vec{C}^{X,C}$ must be made with care regarding the Galilean transformation symmetries (Eqs. (5),(6)). If the original AOEP is already GC, then one expects the constrained $v_{X,C}^\sigma$ to be GC also. Therefore, assuming $w_{X,C}^\sigma(\mathbf{r})$ are GC, the potentials $v_{X,C}^\sigma(\mathbf{r})$ should transform like $v_{X,C}$ in Eqs. (5),(6), under translations and rotations. Therefore, instead of taking the constraints as Eqs. (7)-(9), we construct equivalent constraints which ensure GC and also satisfy Eqs. (7)-(9):

$$\begin{aligned}
\vec{C}^{X,C} &\equiv \left\{ \mathbf{c}_1^{X,C}, \mathbf{c}_2^{X,C}, c_3^{X,C} \right\}, \\
\mathbf{c}_1^{X,C}[n] &= \mathbf{f}_{X,C}[n], \\
\mathbf{c}_2^{X,C}[n] &= \mathbf{t}_{X,C}[n] - \mathbf{D}[n] \times \mathbf{f}_{X,C}[n], \\
c_3^{X,C}[n] &= \Pi_{X,C}[n] + \\
&\quad \int dr \, n(\mathbf{r}) \mathbf{r} \cdot \nabla v_{X,C} - \mathbf{D}[n] \cdot \mathbf{f}_{X,C}[n],
\end{aligned} \qquad (14)$$

where $v_{X,C} \equiv \sum_\sigma v_{X,C}^\sigma$,

$$\begin{aligned}\Pi_X[n] &\equiv E_X[n], \\ \Pi_C[n] &\equiv E_C[n] + T[n] - T_S[n],\end{aligned} \quad (15)$$

and $\mathbf{D}$ is the CM position of the electrons:

$$\begin{aligned}\mathbf{D}[n] &\equiv \frac{1}{N_e} \int dr n(\mathbf{r})\mathbf{r}, \\ N_e &= \int dr n(\mathbf{r}),\end{aligned} \quad (16)$$

It is clear that at the *stationary point*, conditions (14) are equivalent to conditions (5)-(9), since constraints $\mathbf{c}_1^{X,C}$ guarantee zero net force: $\mathbf{f}_{X,C} = \mathbf{0}$. The resulting $v_{X,C}^\sigma$, according to Eqs. (12),(14), are therefore:

$$v_{X,C}^\sigma(\mathbf{r}) = w_{X,C}^\sigma(\mathbf{r}) + \vec{A}^{X,C} \cdot \vec{\Gamma}^\sigma(\mathbf{r}) \quad (17)$$

where

$$\begin{aligned}\vec{A}^{X,C} &\equiv \{\mathbf{a}_1^{X,C}(\mathbf{r}), \mathbf{a}_2^{X,C}(\mathbf{r}), a_3^{X,C}\}, \\ \vec{\Gamma}^\sigma(\mathbf{r}) &\equiv \{\mathbf{g}_1^\sigma(\mathbf{r}), \mathbf{g}_2^\sigma(\mathbf{r}), g_3^\sigma(\mathbf{r})\}, \\ \mathbf{g}_1^\sigma(\mathbf{r}) &\equiv \nabla n_\sigma(\mathbf{r}), \\ \mathbf{g}_2^\sigma(\mathbf{r}) &\equiv \nabla \times [(\mathbf{r}-\mathbf{D})n_\sigma(\mathbf{r})], \\ g_3^\sigma(\mathbf{r}) &\equiv \nabla \cdot [(\mathbf{r}-\mathbf{D})n_\sigma(\mathbf{r})].\end{aligned} \quad (18)$$

The Lagrange multipliers $\{A_i^{X,C}\}$ are determined from the following linear algebraic equations, denoted as

$$\vec{\vec{G}}^{X,C} \cdot \vec{A}^{X,C} = \vec{B}^{X,C}, \quad (19)$$

where

$$\begin{aligned}G_{ij}^{X,C} &= \sum_\sigma \int dr \Gamma_i^\sigma(\mathbf{r}) \Gamma_j^\sigma(\mathbf{r}) \quad \{i,j=1..2d+2\}, \\ \vec{B}^{X,C} &\equiv \Big\{\sum_\sigma \int dr \mathbf{g}_1^\sigma w_{X,C}^\sigma, \sum_\sigma \int dr \mathbf{g}_2^\sigma w_{X,C}^\sigma, \\ &\quad \sum_\sigma \int dr g_3^\sigma w_{X,C}^\sigma + \Pi_{X,C}\Big\}\end{aligned} \quad (20)$$

($d$ is the spatial dimension of the system)

In (20) we used integration by parts to achieve a redefinition of the exact conditions (14), reflected in $\vec{B}^{X,C}$. The fact that the matrices $\vec{\vec{G}}^{X,C}$ are symmetric positive-definite (unless the density is constant) ensures nonsingular solution for $\vec{A}$. It is also possible to constrain only some of the conditions. In this case only the relevant $\{A_i^{X,C}\}$, $\{B_i^{X,C}\}$ and $\{G_{i,j}^{X,C}\}$ are accounted in Eqs.(17) and (19). The last step in our analysis is verifying the GC of $v_{X,C}^\sigma(\mathbf{r})$ in Eq. (17); we refer the reader to appendix A for a proof.

We would like to note here that the global form of the constraints we are imposing, e.g. integrals over all space, may have a size-consistency problem. Thus while approximate AOEP's are size-consistent in the sense that well-separated subsystems have zero interaction, the constrained AOEP's may violate size-consistency because satisfying global constraints for a system of non-interacting components can induce an interaction between them. It is, however, in principle possible to construct *local* constraints when the XC tensor is available.[27] In this case the Lagrange multipliers are calculated locally, at each spatial point and size-consistency is satisfied. Since this method will be far more numerically expensive than enforcing global constraints, we shall not consider it further here..

In summary, the SCF algorithm works as follows. (i) guess initial orbitals $\{\varphi_i^{\sigma 0}\}$, (ii) construct the potentials $v_H$ and $w_{X,C}^\sigma$, (iii) obtain the constrained potentials $v_{X,C}^\sigma$ as described above, (iv) solve KS equations (3), (v) if the stop condition is not satisfied, go back to (ii).

## C. NUMERICAL APPLICATION

We study the effects of the constrained potential for simple $H_2$ chains which nonetheless can be strongly correlated systems. Furthermore, there are some detailed OEP, AOEP and HF results available in the literature, and we use them as a benchmark for our calculations.[28] These calculations refer to exact-exchange (only) KLI (xKLI) and OEP, and HF (the correlation energy is omitted). The xKLI potential has the form[10]

$$w_{xKLI}^\sigma(\mathbf{r}) = \sum_{i=1}^{N} n_i^\sigma(\mathbf{r})\big[u_i^\sigma(\mathbf{r}) + \big(\overline{w}_i^\sigma - \overline{u}_i^\sigma\big)\big]\big/n_\sigma(\mathbf{r}), \quad (21)$$

where

$$u_i^\sigma(\mathbf{r}) \equiv -\frac{1}{\varphi_i^\sigma(\mathbf{r})}\sum_j \varphi_j^\sigma(\mathbf{r}) \int dr' \frac{\varphi_j^{*\sigma}(\mathbf{r}')\varphi_i^\sigma(\mathbf{r}')}{|\mathbf{r}-\mathbf{r}'|},$$
$$n_i^\sigma(\mathbf{r}) \equiv |\varphi_i^\sigma(\mathbf{r})|^2, \quad n^\sigma(\mathbf{r}) \equiv \sum_i |\varphi_i^\sigma(\mathbf{r})|^2,$$
$$\overline{w}_i^\sigma = \langle \varphi_i^\sigma | w_X | \varphi_i^\sigma \rangle, \quad \overline{u}_i^\sigma = \langle \varphi_i^\sigma | u_i | \varphi_i^\sigma \rangle,$$

and $\overline{w}_N^\sigma = \overline{u}_N^\sigma$. The implicit Eq. (21) is solved easily for $\{\overline{w}_i^\sigma\}$, projecting on both sides with each orbital $\varphi_i$, as is described elsewhere.[10]

The relevant exact conditions for this case are for the *exchange* potential, denoted by subscript 'X' in Eqs. (5)-(8). Following Ref.[29] we calculate the first and second polarizability for four potentials; HF, xKLI, xKLI constrained to preserve ZF only ('xKLI-ZF'), and xKLI constrained to preserve both ZF and the Perdew-Levy VT ('xKLI-ZFV') condition. Accordingly, for each such xKLI model, different multipliers $\{A_i^X\}$ are introduced, e.g. for xKLI-ZF we take $\vec{A}^X = \{\mathbf{a}_1^X, \mathbf{0}, \mathbf{0}\}$ in Eqs. (17) and (19), where the only corresponding part of the linear equations (19) are solved: $\sum_j G_{i,j}^X A_j^X = B_j^X, \ i,j = 1..3$. The constrained potential is therefore $v_X^\sigma(\mathbf{r}) = w_{xKLI}^\sigma(\mathbf{r}) + \mathbf{a}_1^X \cdot \mathbf{g}_1^\sigma(\mathbf{r})$. Because of the high symmetry of the H-chains, torque is preserved automatically, in the GS, in accordance with preserving ZF, if the external field acts along the chain.

The specific details of our calculations are as follows. For each chain, the $H_2$ bond length is taken as $2a_0$ ($a_0$ is the Bohr radius) and the distance between the $H_2$ molecules is taken as $3a_0$. Calculations were carried out using our real-space-plane-wave basis set code, with a local pseudopotential for hydrogen.[30] The system was enclosed in a Cartesian box $(L_m + 20) \times 20 \times 20 \ a_0^3$, where $L_m$ is the given chain length. The box is discretized equally along each axis; $\Delta x = \Delta y = \Delta z = 0.364 a_0$. For each chain and potential we calculated the dipole moment of the GS, while applying a dipole electric field along the chain, $v_{ext} = \varepsilon_i x$. From the inversion symmetry of the chains (at $\mathbf{r} = \mathbf{0}$), the polarization can be expanded as a function of odd powers of the applied electric field:

$$\mathbf{P}(\vec{\varepsilon}) = \alpha_{ij} \varepsilon_j + \frac{\gamma_{ijkl}}{3!} \varepsilon_j \varepsilon_k \varepsilon_l + ... \qquad (22)$$

where $\alpha_{ij}$ is the first, or linear polarizability, and $\gamma_{ijkl}$ is the second polarizability. We calculate $\alpha_{xx}$ and $\gamma_{xxxx}$ following method 2 of Ref.[29]: we first calculate the polarizations, $P_x^j$, resulting from a series of electric fields, $\varepsilon_j = 0.002j$, where $j = 0..8$, and use this data to fit the polarizability coefficients using Eq. (22) via least squares. The results, summarized in Table I and TABLE II, are similar to the reported results in Ref.[29], for HF and xKLI. For xKLI-ZF and xKLI-ZFV, the constraints lead to overestimates of the linear polarizability by up to $\sim 1\%$, and up to $\sim 5\%$ for the hyperpolarizability, with respect to regular xKLI. Also, the total energy and eigenvalues for xKLI-ZF and xKLI-ZFV, are very close to xKLI. For instance, the total energy for $H_{12}$, with $v_{ext} = 0.02x$, is about $-175 eV$, and the HOMO energy is about $-7.6 eV$.. The differences between xKLI and xKLI-ZFV are $\sim 2 meV$ for the total energy and $\sim 5 meV$ for the HOMO.

TABLE I. First polarizability $\alpha$ in a.u. for hydrogen chains.

|          | HF    | xKLI  | xKLI-ZF | xKLI-ZFV |
|----------|-------|-------|---------|----------|
| $H_4$    | 32.2  | 33.3  | 33.6    | 33.6     |
| $H_6$    | 56.7  | 60.6  | 61.3    | 61.3     |
| $H_8$    | 83.8  | 91.8  | 93.1    | 93.1     |
| $H_{12}$ | 140.1 | 159.4 | 161.9   | 161.9    |

TABLE II. Second polarizability $\gamma/10^3$ in a.u. for hydrogen chains

|          | HF    | xKLI  | xKLI-ZF | xKLI-ZFV |
|----------|-------|-------|---------|----------|
| $H_4$    | 10.4  | 10.5  | 10.8    | 10.8     |
| $H_6$    | 29.7  | 35.4  | 36.7    | 36.7     |
| $H_8$    | 61.8  | 90.4  | 94.0    | 93.9     |
| $H_{12}$ | 152.5 | 304.5 | 317.6   | 318.1    |

The same trend is found for some small sodium clusters, where the xKLI-ZF/xKLI-ZFV values for the GS energy and eigenvalues are negligibly different from xKLI. The first polarizability tensor, $\vec{\vec{\alpha}}$, is slightly overestimated by the xKLI-ZF and xKLI-ZFV with respect to the xKLI results.

The small overestimation of the polarizabilities in the results where exact conditions are enforced suggests that a very small amount of self-interaction has been introduced, due to the additional local density terms $\vec{\Gamma}$, in the constrained potential $v^\sigma_{X,C}$ (Eqs. (15),(16)).

## III. EXTENSION TO TIME DEPENDENT XC POTENTIALS

### A. EXACT CONDITIONS IN TIME-DEPENDENT DENSITY FUNCTIONAL THEORY

While derivation of exact conditions in DFT resulted from a variation of the energy *functional*, in the time-dependent theory things are more complicated.[4] The present lack of a real-time action functional requires alternative formalisms such as the Keldysh contour method.[24, 31-33] However, some exact conditions may be formulated without such an action, and are actually TD versions of the DFT conditions (see Sec. IIB). We refer to four of them: (i) generalized translational covariance of the $v_{XC}(\mathbf{r},t) \equiv v_X(\mathbf{r},t) + v_C(\mathbf{r},t)$. i.e. for any *TD* translation $\mathbf{r}' = \mathbf{r} + \mathbf{x}(t)$:[34]

$$v_{XC}[n'](\mathbf{r},t) = v_{XC}[n](\mathbf{r}+\mathbf{x}(t),t). \qquad (23)$$

This condition results from the harmonic potential theorem.[35] (ii) zero net XC-force,[34]

$$\mathbf{f}_{XC}(t) \equiv -\sum_\sigma \int d\mathbf{r}\, n_\sigma(\mathbf{r},t) \nabla v^\sigma_{XC}(\mathbf{r},t) = \mathbf{0}, \qquad (24)$$

which results from Ehrenfest's equations.

For adiabatic energy functionals, like the exact exchange, a TD energy functional can be defined as $E(t) \equiv E[\{\varphi^\sigma_i(\mathbf{r},t)\}]$. The functional $E[\{\varphi^\sigma_i(\mathbf{r},t)\}]$ is the time-independent energy functional defined in Eq. (1). The other functionals in (1) can be defined correspondingly, e.g. $E_{XC}(t) \equiv E_{XC}[\{\varphi^\sigma_i(\mathbf{r},t)\}]$. In such a case, there are two more conditions: (iii) the virial theorem[36]

$$\begin{aligned}E_{XC}[n] = &-\sum_\sigma \int d\mathbf{r}\, n_\sigma(\mathbf{r},t)\,\mathbf{r}\cdot\nabla v^\sigma_{XC}(\mathbf{r},t) + \\ &T_S[n](t) - T[n](t),\end{aligned} \qquad (25)$$

and (iv) the XC energy balance (EB)[36]

$$\dot{E}_{XC}(t) = \sum_\sigma \int d\mathbf{r}\, v^\sigma_{XC}(\mathbf{r},t)\,\dot{n}_\sigma(\mathbf{r},t), \qquad (26)$$

which is required from the energy balance of the interacting TD energy, $E(t)$, and the non-interacting energy, $E_S(t)$:

$$\begin{aligned}\dot{E}(t) &= \dot{W}_{ext}, \\ \dot{E}_S(t) &= \dot{W}_S,\end{aligned} \qquad (27)$$

where $W_{ext,S} \equiv \int_0^t dt' \int d\mathbf{r}\, \dot{v}_{ext,S}(\mathbf{r},t')\, n(\mathbf{r},t')$ is the total external mechanical work done by the external potential $v_{ext,S}$.

Note that conditions (23)-(25) are formulated for the total XC potential rather than for the exchange and correlation potentials separately.

## B. IMPOSING EXACT CONDITIONS ON A GIVEN TIME-DEPENDENT XC POTENTIAL

Imposing the exact conditions in the TD case is done here similarly to the time-independent case, described in sec. IIB. Here, however, we search for *each* time $t$ and given TD orbitals (and $w_{XC}^\sigma(\mathbf{r},t)$ ), for the closest $v_{XC}^\sigma(\mathbf{r},t)$ to $w_{XC}^\sigma(\mathbf{r},t)$.

We formulate, again, a variational problem for the constrained TD Lagrangian

$$\ell(t) = \tfrac{1}{2}\sum_\sigma \int dr \left| v_{XC}^\sigma(\mathbf{r},t) - w_{XC}^\sigma(\mathbf{r},t) \right|^2 + \vec{A}(t)\cdot\vec{C}(t). \quad (28)$$

Correspondingly, we construct the relevant TD constraints, $\mathbf{C}(t) \equiv \{\mathbf{c}_1(t), c_2(t), c_3(t)\}$, by replacing $n(\mathbf{r})$ by $n(\mathbf{r},t)$ for the relevant physical quantities in Eqs. (10)-(15), i.e.

$$\begin{aligned}
\mathbf{c}_1(t) &= \mathbf{c}_1^X[n(\mathbf{r},t)] + \mathbf{c}_1^C[n(\mathbf{r},t)], \\
c_2(t) &= \dot{E}_{XC}(t) - \sum_\sigma \int dr v_{XC}^\sigma(\mathbf{r},t) \dot{N}_\sigma(\mathbf{r},t), \\
c_3(t) &= c_3^X[n(\mathbf{r},t)] + c_3^C[n(\mathbf{r},t)],
\end{aligned} \quad (29)$$

where

$$\begin{aligned}
\dot{N}_\sigma(\mathbf{r},t) &\equiv [\dot{n}_\sigma(\mathbf{r},t) + \nabla n_\sigma(\mathbf{r},t)\cdot \dot{\mathbf{D}}(t)] = \\
&-\nabla \cdot [\mathbf{j}_\sigma(\mathbf{r},t) - n_\sigma(\mathbf{r},t)\dot{\mathbf{D}}(t)]
\end{aligned}$$

$v_{XC}(\mathbf{r},t) = \sum_\sigma v_{XC}^\sigma(\mathbf{r},t)$, $\{c_i^{X,C}[n]\}$ are defined in Eq. (14), and $\mathbf{D}(t) \equiv \mathbf{D}[n(\mathbf{r},t)]$ is now the TD CM position of the electron density.

It is clear, again, that the conditions in (29), at the stationary point, are equivalent to the conditions (24)-(26), since $\mathbf{f}_{XC}(t) = \mathbf{0}$ is enforced by $\mathbf{c}_1(t)$. Hence, following the variation of $\ell(t)$, the corrected potential is (similar to Eq. (17))

$$v_{XC}^\sigma(\mathbf{r},t) = w_{XC}^\sigma(\mathbf{r},t) + \vec{A}(t)\cdot\vec{\Lambda}^\sigma(\mathbf{r},t), \quad (30)$$

where

$$\begin{aligned}
\vec{\Lambda} &\equiv \{\mathbf{l}_1^\sigma, l_2^\sigma, l_3^\sigma\}, \\
\mathbf{l}_1^\sigma(\mathbf{r},t) &\equiv \nabla n_\sigma(\mathbf{r},t), \\
l_2^\sigma(\mathbf{r},t) &\equiv \dot{N}_\sigma(\mathbf{r},t), \\
l_3^\sigma(\mathbf{r},t) &\equiv \nabla\cdot[(\mathbf{r}-\mathbf{D}(t))n_\sigma(\mathbf{r},t)].
\end{aligned} \quad (31)$$

The TD Lagrange multipliers are determined from the linear algebraic equations,

$$\vec{\vec{L}}(t)\cdot\vec{A}(t) = \vec{B}(t), \quad (32)$$

which must be solved separately for each time, $t$, where:

$$\begin{aligned}
L_{ij}(t) &= \sum_\sigma \int dr\, l_i^\sigma(\mathbf{r},t) l_j^\sigma(\mathbf{r},t) \quad \{i,j = 1..d+2\}, \\
\vec{A}(t) &\equiv \{\mathbf{a}_1, a_2, a_3\}, \\
\vec{B}(t) &\equiv \Big\{ \sum_\sigma \int d\mathbf{r}\, \mathbf{l}_1^\sigma w_{XC}^\sigma, \dot{E}_{XC} - \sum_\sigma \int dr\, l_2^\sigma w_{XC}^\sigma, \\
&\quad \sum_\sigma \int dr\, l_3^\sigma w_{XC}^\sigma + \Pi_X + \Pi_C \Big\}.
\end{aligned} \quad (33)$$

($d$ is the spatial dimension of the system).

The fact that the matrix $\vec{\vec{L}}$ is symmetric and positive-definite for $t > 0$, ensures a unique solution for $\vec{A}$. Also note that the TD and time-independent potentials in (30) and (17), respectively, should be matched at $t = 0$, in order to guarantee continuity in time. Since TD and time-independent conditions are not exactly the same, a proper match is desired at $t = 0$. For instance, it is possible to replace $w_{XC}^\sigma(\mathbf{r},t)$ in Eqs. (28), (29), (30) and (33) by $\bar{w}_{XC}^\sigma(\mathbf{r},t) = w_{XC}^\sigma(\mathbf{r},t) + \sum_i A_i^0 \Gamma_i^0(\mathbf{r},t)$, in order to achieve continuity across $t = 0$. In the last equality, $\{A_i^0\}$ are the time-independent Lagrange multipliers, calculated for the GS, and the $\Gamma_i$'s of Eq.(18) are substituted. Another possibility for achieving continuity across $t = 0$, is by enforcing all the GS conditions plus the EB condition from Eq. (26) (despite the fact that zero net torque is not necessary for the TD XC potential). Finally, we refer the reader to Appendix A to verify that $v_{XC}^\sigma(\mathbf{r},t)$ in Eq. (30) is TC.

## C. NUMERICAL IMPLEMENTATION

The numerical procedure of propagating the TDKS equations requires calculating the GS first. At this stage the SCF is obtained as is described in Sec. IIB. Then, the orbitals are propagated in time by the TDKS equations:[5]

$$\begin{aligned}
i\dot{\varphi}_i^\sigma(\mathbf{r},t) &= \hat{h}_S^\sigma(\mathbf{r},t)\varphi_i^\sigma(\mathbf{r},t), \\
\hat{h}_S^\sigma(\mathbf{r},t) &= -\nabla^2/2 + v_{ext}(\mathbf{r},t) + \\
&\quad v_H[n](\mathbf{r},t) + v_{XC}^\sigma[\{\varphi_i^\sigma\}](\mathbf{r},t),
\end{aligned} \quad (34)$$

as follows. For each time step, first $w_{XC}^\sigma(\mathbf{r},t)$ is constructed from the actual KS orbitals, then, $v_{XC}^\sigma(\mathbf{r},t)$ is calculated as is described in IIIB, and finally, $\{\varphi_i^\sigma(\mathbf{r},t+\Delta t)\}$ are obtained from Eqs. (34).

For the case that ionization processes are occurring, an imaginary absorbing potential is the most convenient way to treat the ionized electrons in our code, which is operates in real time and space. In this case, we add the following negative imaginary potential[37]

$$v_{abs}(\mathbf{r}) = -iu_{abs}(\mathbf{r}),$$
$$u_{abs}(\mathbf{r}) = \eta \times \theta(|x| - a) \times (|x| - a)^n, \qquad (35)$$

The parameters $\eta$, $n$, are taken from Ref. [37], $a$ is chosen as the plane $x = a$ which is far away from the GS CM, and $\theta(x)$ is the step function. In Eq. (35) we assume that the driving field and therefore any ionization process acts along the $x$ axis. The Hamiltonian $\hat{h}_S^\sigma$ in Eq. (34) is this case is replaced by $\hat{\tilde{h}}_S^\sigma \equiv \hat{h}_S^\sigma + v_{abs}$. Correspondingly, the evolution of TD expectation values are changed, where for each Hermitian operator $\hat{o}$ we can write

$$d\langle \tilde{\hat{o}} \rangle / dt \equiv i\langle [\hat{h}_S, \hat{o}] \rangle + \langle \partial \hat{o} / \partial t \rangle - \langle u_{abs}\hat{o} + \hat{o}u_{abs} \rangle. \qquad (36)$$

Therefore, the continuity equation, the total number of electrons, the CM position and the mechanical work are changed to

$$\tilde{\dot{n}}(\mathbf{r},t) = -\nabla \cdot \mathbf{j}(\mathbf{r},t) - 2n(\mathbf{r},t)u_{abs}(\mathbf{r}),$$
$$\tilde{\dot{N}}_e(t) = -2\int dr\, n(\mathbf{r},t)u_{abs}(\mathbf{r}),$$
$$\tilde{\dot{\mathbf{D}}}(t) \equiv \dot{\mathbf{D}}(t) - 2\int dr\, n(\mathbf{r},t)u_{abs}(\mathbf{r})\mathbf{r}/N_e(t), \qquad (37)$$
$$\tilde{\dot{W}}_{ext}(t) \equiv \dot{W}_{ext}(t) - 2\int dr\, u_{abs}(\mathbf{r})v_S(\mathbf{r},t)n(\mathbf{r},t) -$$
$$\mathrm{Re}\sum_{i,\sigma}\int dr\, \varphi_i^\sigma(\mathbf{r},t)^* u_{abs}(\mathbf{r})\nabla^2\varphi_i^\sigma(\mathbf{r},t),$$

where $\dot{\mathbf{D}} = -\left(\int dr\, \nabla \cdot \mathbf{j}(\mathbf{r},t)\mathbf{r}/N_e + \mathbf{D} \times \dot{N}_e/N_e\right)$. Also, note that $\dot{N}_\sigma(\mathbf{r},t)$ (Eq. (29)) and $\dot{E}_{XC}$ (Eq. (33)) should be changed in correspondence to Eq. (37) and (36), respectively.

An important numerical issue is connected with the inherent singularity of the Lagrange multiplier $A_{d+1}(t)$, related to the EB condition. In the linear response regime, the deviation from the exact EB is $\Delta \dot{E}_{XC,1} \approx \int dr\left[v_{XC}^{exact}(\mathbf{r}) - w_{XC}(\mathbf{r})\right]\dot{n}_1(\mathbf{r},t)$. Looking at the left and right hand sides of the corresponding linear equation, $\sum_i L_{d+1,i}(t)A_i(t) = B_{d+1}(t)$, we can see that

$$L_{d+1,1..d},\ L_{d+1,d+2},\ B_{d+1} \sim o(\dot{n}_1),$$
$$L_{d+1,d+1} = \sum_\sigma \int dr\, l_2^\sigma(\mathbf{r},t)^2 \sim o(\dot{n}_1^2).$$

Therefore, $A_{d+1}(t) \sim o(1/\dot{n}_1)$. The consequence of this divergence for $|\dot{n}_1| \to 0$ is the need for significantly smaller time-steps, down to $10^{-4} - 10^{-5}$ a.u., which makes calculations impractical. Furthermore, negative values of $A_{d+1}$ may yield 'hydrodynamical' instabilities, while analyzing the quantum equation of motion for the current density (see Appendix B). Therefore, in order to overcome this singularity and other potential numerical instabilities, we calculate $\vec{A}$ from a set of regularized linear equations, as is described in appendix B. As a result, the EB condition will be preserved on the average, approximately, while the other conditions will be preserved exactly.

## D.  APPLICATIONS TO ELECTRON DYNAMICS IN SODIUM CLUSTERS

For the TD case, we test our constraining method on electron dynamics in small sodium clusters. We tested the performance of $v_{XC}^\sigma(\mathbf{r},t)$, where $w_{XC}^\sigma(\mathbf{r},t)$ is taken as the TD version of the xKLI (TD-xKLI) potential, as defined by substituting the TD orbitals $\{\varphi_i^\sigma(\mathbf{r},t)\}$ in the $w_{xKLI}^\sigma$ of Eq. (21). We mainly focus on the case of $Na_5$, where pathological behavior had already reported by Mundt and Kummel (MK), in Ref. [22]. MK found that for weak initial perturbation of $Na_5$ (initially in the GS) treated by xKLI, the TD dipole moment is unphysically amplified, corresponding to ZF violation. We do confirm their observation for the case of a short weak driving field. Moreover, even when we enforce ZF, ZT, and the VT during the propagation, the $Na_5$ dipole moment is also unphysically amplified, after some tens of fs,. Furthermore, enforcing both EB and ZF does not yield better results, since EB is only preserved approximately, due to the singularity discussed above (see IIIC). On the other hand, for other clusters like $Na_4$, $Na_5^+$, and $Na_6$,[38] we do not face such unphysical excitations (despite the ZF and EB violations).

Our method, despite its inefficiency for finite time-interval driving field, is applicable when (either weak or strong) CW field is applied. In this case, the CW field generates an oscillating $\dot{N}(\mathbf{r},t)^2$ field, where $\dot{N}(\mathbf{r},t)^2 \to 0$ for short periods. During such periods, our regularization prevents catastrophic timestep reductions, and in turn, the EB condition is recovered at other nonsingular time intervals.

We present some typical results of electron dynamics in the model $Na_5$ cluster. The cluster geometry is a planar $(x-y)$ trapezoid,[39] where the bases are parallel to the x-axis. Also, we slightly broke symmetry within the x-y plane, to make the problem more general. The sodium atoms are located at the following x-y coordinates: $(7.36,-2.9)$, $(-7.06,-2.9)$, $(-0.1,-2.9)$, $(3.6,4)$ and $(-3.8,4.7)$. We used a local pseudopotential for the s-electron of each sodium atom, described in Ref. [38].

Following the SCF calculation, we apply a CW laser field:

$$v_{laser}(\mathbf{r},t) = -x \times E_0 \sin^2(\pi/T_p)\cos(\omega_{opt}t), \qquad (38)$$

where $\omega_{opt} \equiv 2\pi/T_{opt}$ is the optical laser frequency, $T_p$ is the laser pulse duration, and $E_0$ is the amplitude. For a given $v_{laser}$ we compare four calculations: TDHF, and three TD-xKLI models having the general form

$$\begin{aligned}v_{XC}^\sigma(\mathbf{r},t) = \\ w_{xKLI}^\sigma(\mathbf{r},t) + \mathbf{a}_1(t)\cdot\nabla n_\sigma(\mathbf{r},t) + a_2(t)\dot{N}_\sigma(\mathbf{r},t),\end{aligned} \qquad (39)$$

For TD-xKLI without any constraints: $\mathbf{a}_1(t) \equiv \mathbf{0}$, $a_2(t) \equiv 0$. For TD-xKLI-ZF: $a_2(t) \equiv 0$, and for TD-xKLI-ZF-EB we fit both $\mathbf{a}_1$ and $a_2$. Recall that since ionization is involved we include the absorbing potential, Eq. (35), in the TDKS Hamiltonian. Therefore, the tilda quantities, defined in Eq. (37), are substituted in Eq.(39). Also, for consistency (and simplicity) we enforced exactly the same constraints at the SCF stage for each model (e.g. for TD-xKLI-ZF/TD-xKLI-ZF-EB, only ZF is enforced in the SCF).

As diagnostics, we record the TD dipole moment $\mathbf{d}(t) \equiv \int dr n(\mathbf{r},t) \mathbf{r}$, the total number of ionized electrons, $N_I(t) \equiv N_e(0) - N_e(t)$, the energy conservation parameter $\Delta E_B(t) \equiv E(t) - \tilde{W}_{ext}(t)$, and the moment of inertia tensor, $\{I(t)\}_{ij} = \int dr n(\mathbf{r},t)[r^2 \delta_{ij} - r_i r_j]$. Finally, at the end of the time propagation we also calculate the emission power spectra of $d_x(t)$: $P_x(\omega) \equiv |\omega^2 \tilde{d}_x(\omega)|^2$, where $\tilde{d}_x(\omega)$ is the Fourier transform of $d_x(t)$. We performed all calculations on a $75 \times 45 \times 45 \, a_0^3$ Cartesian grid, where a uniform mesh spacing, $\Delta x = \Delta y = \Delta z = 1 a_0$, was used. We used optimized parameter values, $n = 3$, $\eta = 10^{-4}$ and $a = 17 a_0$ for the absorbing potential in Eq. (35).

We present here calculations for three different laser pulses (LP), summarized in Table 3.

Table III. The properties of 3 laser pulses used here (see Eq. (38)).

| Label | $E_0 \, (\text{H}/a_0)$ | $\omega_{opt}$ (a.u.) | $T_p/T_{opt}$ | $I_{laser} \, (\text{W/cm}^2)$ |
|---|---|---|---|---|
| LP1 | 0.001 | 0.05 | $15 \, (\sim 46\text{fs})$ | $3.5 \times 10^{10}$ |
| LP2 | 0.01 | 0.05 | $16 \, (\sim 49\text{fs})$ | $3.5 \times 10^{12}$ |
| LP3 | 0.01 | 0.04 | $13 \, (\sim 50\text{fs})$ | $3.5 \times 10^{12}$ |

The results for the energy conservation parameter, $\Delta E_B(t)$, are depicted in FIG. 1, for each laser pulse shown in Table III. For LP1 (FIG. 1a) and LP2 (FIG. 1b), we see a consistent hierarchy, where xKLI-ZF-EB is the most energy conserving model and xKLI is least conserving. In FIG. 1c, on the other hand, we notice that the xKLI-ZF is the worst model for energy conservation. For this case, we checked numerical convergence with respect to both time step size and spatial meshing, resulting in similar results -- we note that $\Delta E_B$ increasing is due to both $\tilde{W}_{ext}$ decreasing and the total energy, $E$, increasing. Typically, xKLI-ZF-EB conserves energy to about a few times $10^{-2}$ a.u., while the other KLI-models are only good to about $0.1$ a.u. or more. On the other hand, the energy conservation of the TDHF method is 3-4 orders of magnitude better than xKLI-ZF-EB. The characteristic total work, $\tilde{W}_{ext}(T_p)$, is $\sim 0.02 - 0.05$ a.u. for LP1, $\sim 1.2 - 1.5$ a.u. for LP2 and $\sim 0.22 - 0.8$ a.u. for LP3, where $0.22$ a.u. refers to TD-xKLI-ZF, while the rest are $\sim 0.7 - 0.8$ a.u.

In FIG. 2 we compare the total number of ionized electrons for each model, resulting from the LP2 and LP3 pulses. Here, again, xKLI-ZF-EB results are the KLI model that is closest to TDHF. It is also interesting to see, again, that xKLI-ZF is not necessarily a better model for ionization than xKLI, as is shown in FIG. 2a. In FIG. 3 we show the 'transversal dynamics', for LP3, characterized by the moment of inertia eigenvalue related to the x-axis (general) direction. This quantity indicates the electron density distribution, perpendicular to the laser field direction. Here, at the initial stage of the dynamics, up to 700-800 a.u., all KLI models and HF seem quite similar, while at the final stage, xKLI-ZF-EB and HF remain close and seem separated from the other KLI models.

Finally, we compare the emission power spectra for LP2, in FIG. 4, where each xKLI model is compared individually to the HF spectra. Here, xKLI-ZF-EB does not exhibit better results, as it did for the previously discussed quantities. For instance, the first three harmonics in FIG. 4 look similar for all three xKLI models, and it is only for a few harmonics, e.g. the 9$^{th}$ harmonic, that xKLI-ZF-EB looks notably most similar to the HF spectra. Furthermore, for the very high harmonics, not shown here, xKLI-ZF-EB continues to be similar to the other xKLI models (all overestimate the power spectra with respect to HF).

## IV. DISCUSSIONS AND CONCLUSIONS

We have developed a practical method that constrains any given XC potential to preserve some exact conditions. The method is numerically stable and efficient, for both DFT and TDDFT. We tested the performance of constrained xKLI potentials for both time-dependent and independent cases. In the time-independent case, the GS results are negligibly different to the non-conserving xKLI results, for some H-chains and small sodium clusters. We found very small and small overestimations of the first and second polarizabilities, respectively, with respect to xKLI when constraints are satisfied. We explain this overestimation by the introduction of a very small amount of self-interaction via the local density correction, $\vec{A}^X \cdot \vec{\Gamma}^\sigma [n](\mathbf{r})$, in Eq. (17). We conclude that preserving exact conditions does not yield worthwhile improvements for the examples discussed.

On the other hand, our method for ensuring constraints are satisfied is important for the TD case, and does exhibit improved results with respect to TDHF. We demonstrated the importance of preserving *both* ZF and EB during the time-propagation. The xKLI-ZF-EB method preserves energy typically by at least by an order of magnitude better than the other xKLI models. It also least overestimates total ionization relative to TDHF, and has 'transversal dynamics' that is more similar to TDHF than the other xKLI models. On the other hand, the emission power spectrum via xKLI-ZF-EB is not significantly improved relative to xKLI.

The main problem that prevents full energy conservation is likely to be the lack of EB, even in first order, as is explained in Sec. IIIC. To overcome this problem we suggest a possible alternative method, which unfortunately cannot be applied in our numerical code. Since time-independent OEP can be done relatively routinely now,[40] we suggest calculating the GS, first, by the exact OEP, $w_{OEP}(\mathbf{r})$, and then, propagate by the TD-AOEP: $w_{AOEP}(\mathbf{r},t) - w_{AOEP}(\mathbf{r}) + w_{OEP}(\mathbf{r})$ and its constrained versions. It is clear, that in first order, the EB deviation will be $\Delta E_{XC,1}(t) = 0$ (see sec. IIIC). Consequently, the singularity should also disappear, and therefore the constraints can be applied without divergences (see Sec. IIIC).

It should also be reiterated that constraining approximate OEP's to ensure known exact conditions are satisfied is treating symptoms (violation of such conditions) rather than the disease that is causing those symptoms (which is the failure of AOEP's to be a complete functional derivative of an XC energy functional). In this respect, the improvements obtained for time-dependent observables are encouraging, but the underlying limitation remains.

Finally, some insight about the, as yet unexplained, TD-OEP instability[17] may be gleaned from this problem regarding the stability of the EB constraint. Recall that the term $a_2 \dot{N}(\mathbf{r},t)$ in Eq. (39) may yield a 'hydrodynamical' instability. As explained in Appendix B, an explicit current-density dependent XC potential may yield a 'negative viscosity' term in the current-density evolution equation, resulting in in numerical instabilities. Unfortunately, such terms appear in the regular version of TD-OEP,[16] developed by Ullrich, Gossmann and Gross (see Eqs. (18),(19) in Ref. [16]). We therefore suggest checking this issue at least for simple cases.

## ACKNOWLEDGEMENTS

We would like to thank Dr. Leeor Kronik and Prof. Stephan Kummel, for providing us their detailed calculations, regarding hyperpolarizabilities of H-chains. Also, Y.K. gratefully thanks Prof. Roi Baer and his research group, for hosting his numerical code, patiently, on their computers cluster, 'CUBIOT'. This work was supported by the Director, Office of Science, Office of Basic Energy Sciences, of the US Department of Energy under contract DE-AC02-05CH11231 (the Ultrafast X-ray Science Laboratory).

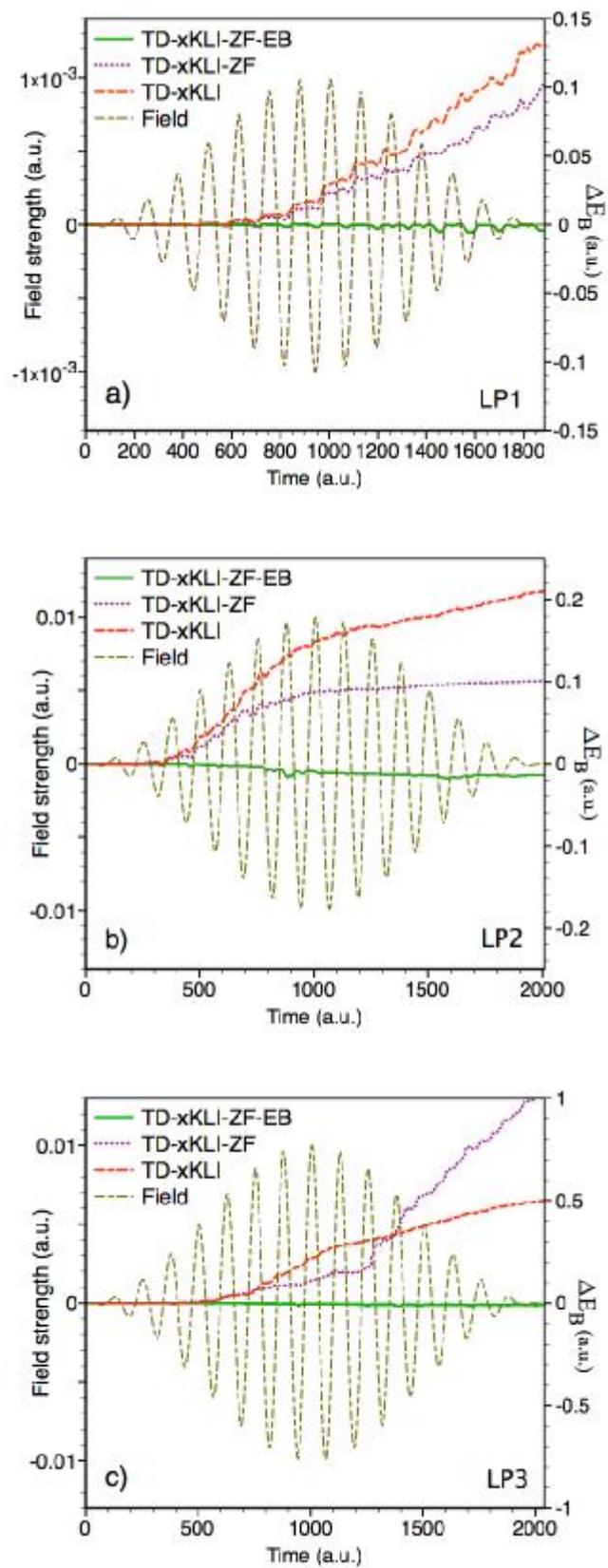

FIG. 1 Energy conservation (right scale) and field strength (left scale) vs. time for the laser pulses: a) LP1, b) LP2, c) LP3.

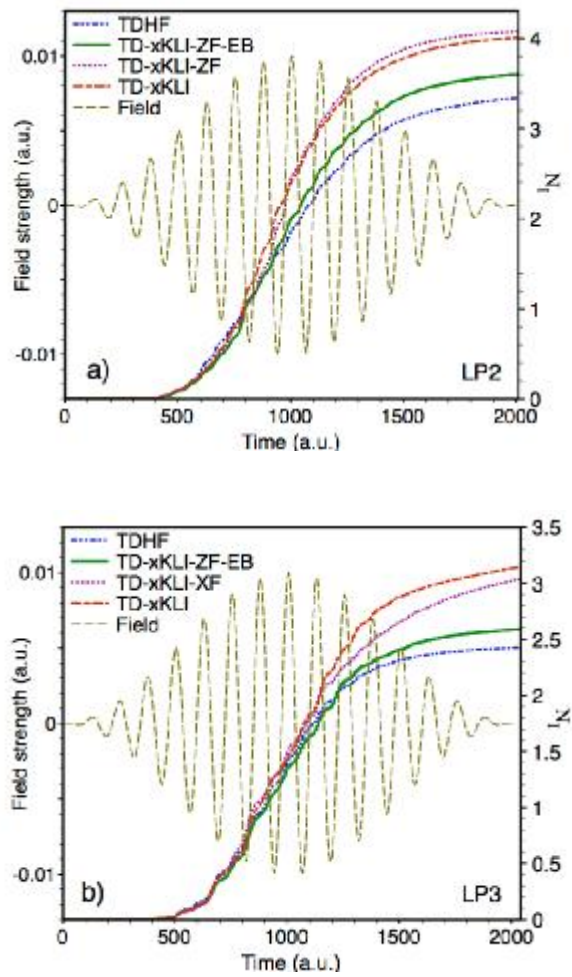

FIG. 2 Total number of ionized electrons (right scale) and field strength (left scale) vs. time for the laser pulses: a) LP2, b) LP3. (Ionization for LP1 is very weak for all models, therefore is not depicted).

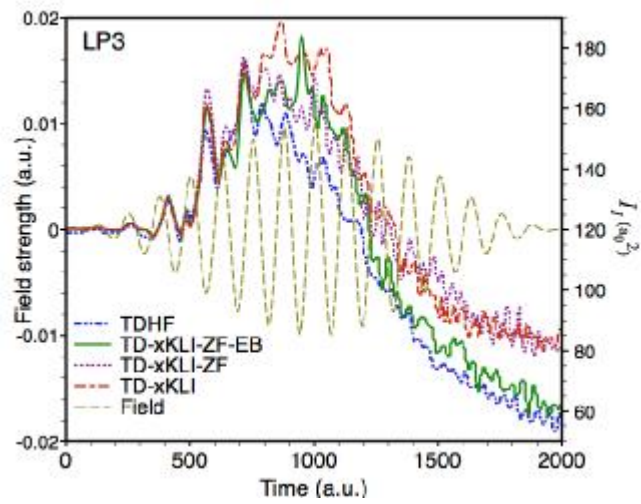

FIG. 3 Transverse dynamics: moment of inertia eigenvalue $I_1\left(a_0^2\right)$ (right scale) and field strength (left scale) vs. time, for laser pulse LP3.

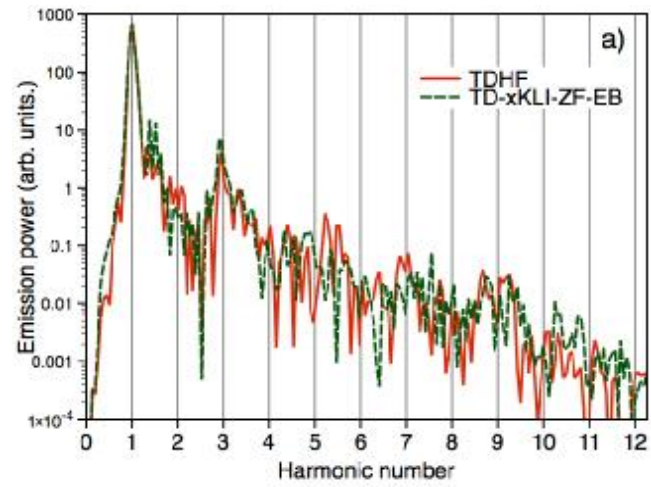

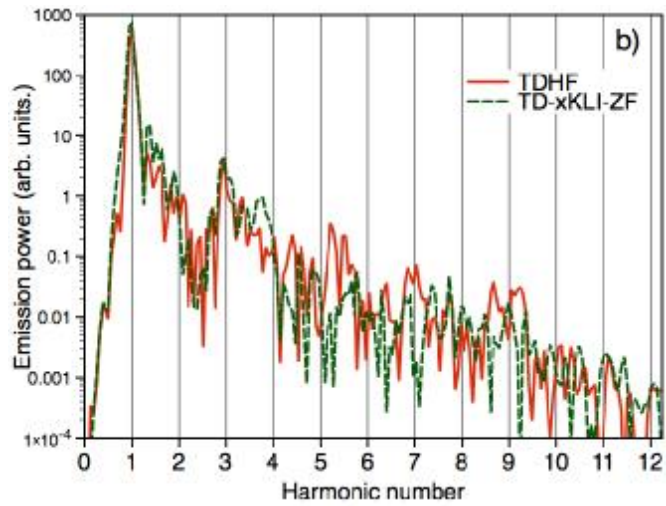

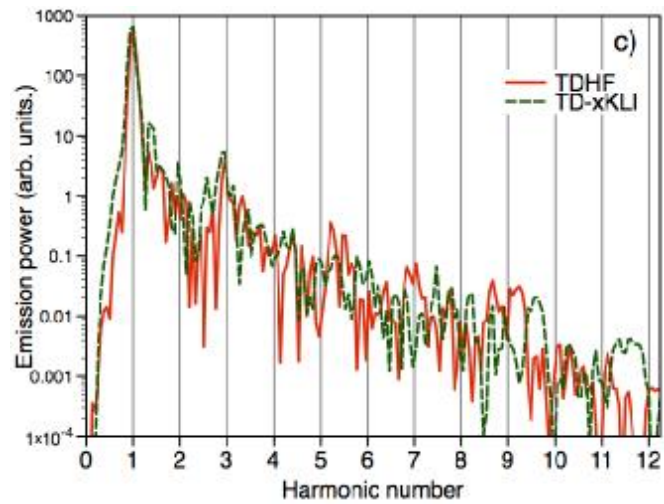

FIG. 4 Emission power spectra (arb. units) vs. harmonic number for laser pulse LP2. Each of the models (green lines) in the Figs. a) TD-xKLI-ZF-EB, b) TD-xKLI-ZF, c) TD-xKLI, is compared to TDHF (red lines) emission power spectra.
20

## APPENDIX A: TRANSLATIONAL AND ROTATIONAL COVARIANCE PROOFS

We prove in this appendix the GC of the time-independent constrained X,C potentials, $v_{X,C}^\sigma$ (Eq. (17)), and the TC of the TD XC potential of Eq. (30).

Consider two fixed Cartesian coordinates systems, $\mathbf{r}'$ and $\mathbf{r}$. Under translation with an arbitrary vector, $\mathbf{x}$, the density is transformed as

$$\begin{aligned} n'_\sigma(\mathbf{r}) &= n_\sigma(\mathbf{r}+\mathbf{x}), \\ n'(\mathbf{r}) &\equiv \sum_\sigma n'_\sigma(\mathbf{r}) = n(\mathbf{r}), \end{aligned} \quad (A1)$$

the CM position transforms as:

$$\begin{aligned} \mathbf{D}[n'] &= \frac{1}{N_e}\int dr\, n'(\mathbf{r})\mathbf{r} = \\ &\frac{1}{N_e}\int dr\, n(\mathbf{r}+\mathbf{x})\mathbf{r} = \mathbf{D}[n] - \mathbf{x}, \end{aligned} \quad (A2)$$

and correspondingly the $\Gamma'_i s$ transform as:

$$\begin{aligned} \mathbf{g}_1^\sigma[n'](\mathbf{r}) &\equiv \nabla n'_\sigma(\mathbf{r}) = \nabla n_\sigma(\mathbf{r}+\mathbf{x}) = \mathbf{g}_1^\sigma[n](\mathbf{r}+\mathbf{x}), \\ \mathbf{g}_2^\sigma[n'](\mathbf{r}) &\equiv \nabla \times [(\mathbf{r}-\mathbf{D}')n'_\sigma(\mathbf{r})] = \\ &\nabla \times [(\mathbf{r}-\mathbf{D}+\mathbf{x})n_\sigma(\mathbf{r}+\mathbf{x})] = \mathbf{g}_2^\sigma[n'](\mathbf{r}+\mathbf{x}), \\ g_3^\sigma[n'](\mathbf{r}) &\equiv \nabla \cdot [(\mathbf{r}-\mathbf{D}')n'_\sigma(\mathbf{r})] = \\ &\nabla \cdot [(\mathbf{r}-\mathbf{D}+\mathbf{x})n_\sigma(\mathbf{r}+\mathbf{x})] = g_3^\sigma[n](\mathbf{r}+\mathbf{x}). \end{aligned} \quad (A3)$$

The matrix $G^{X,C}$ and the vector $\vec{B}^{X,C}$, in Eqs. (20), are constructed as spatial integrations of the $\{\Gamma_i\}$ and the original potential $w_{X,C}^\sigma$. The former is assumed to be TC, therefore,

$$\begin{aligned} \overset{\leftrightarrow}{G}'^{X,C} &= \overset{\leftrightarrow}{G}^{X,C}, \\ \vec{B}'^{X,C} &= \vec{B}^{X,C}, \\ \vec{A}'^{X,C} &= \left(\overset{\leftrightarrow}{G}'^{X,C}\right)^{-1}\vec{B}'^{X,C} = \\ &\left(\overset{\leftrightarrow}{G}^{X,C}\right)^{-1}\vec{B}^{X,C} = \vec{A}^{X,C}, \end{aligned} \quad (A4)$$

and thus

$$\vec{A}'^{X,C} \cdot \vec{\Gamma}'(\mathbf{r}) = \vec{A}^{X,C} \cdot \vec{\Gamma}(\mathbf{r}+\mathbf{x}), \quad (A5)$$

i.e. $v_{XC}^{\sigma}$ is TC.

Hence, under a rigid transformation by $d \times d$ rotation matrix $M$, the density and the CM transform as

$$n'(\mathbf{r}) = n(M\mathbf{r}),$$
$$\mathbf{D}[n'] = \frac{1}{N_e} \int dr n(M\mathbf{r})\mathbf{r} = M^{-1}\mathbf{D}[n]. \quad (A6)$$

Together with the nabla transformation $[\nabla f]_i(\mathbf{r}') = (M^{-1})_{ij}[\nabla f]_j(M\mathbf{r})$, where $f$ is a scalar function, and $[\nabla f]_i(\mathbf{r}) \equiv \partial f(\mathbf{R})/\partial R_i\big|_{\mathbf{R}=\mathbf{r}}$, we find that

$$\mathbf{g}_1^{\sigma}[n'](\mathbf{r}) = M^{-1}\mathbf{g}_1^{\sigma}[n](M\mathbf{r}),$$
$$\mathbf{g}_2^{\sigma}[n'](\mathbf{r}) = M^{-1}\mathbf{g}_2^{\sigma}[n](M\mathbf{r}), \quad (A7)$$
$$g_3^{\sigma}[n'](\mathbf{r}) \equiv g_3^{\sigma}[n](M\mathbf{r}).$$

Hence, the matrix $G^{X,C}$ and the vector $\vec{B}^{X,C}$, transform, by construction again, as

$$\vec{\vec{G}}'^{X,C} = O\vec{\vec{G}}O^T,$$
$$\vec{B}'^{X,C} = O\vec{B}^{X,C}, \quad (A8)$$

where the $(2d+1) \times (2d+1)$ matrix $O$ is

$$O = \begin{pmatrix} M & & \\ & M & \\ & & 1 \end{pmatrix} \quad (A9)$$

(d is the spatial dimensions).

Thus, following Eq. (A8), the equation

$$\vec{\vec{G}}'^{X,C} \cdot \vec{A}'^{X,C} = \vec{B}'^{X,C}, \quad (A10)$$

can be rewritten as

$$O\vec{G}^{X,C}O^T \cdot \vec{A}'^{X,C} = O\vec{B}^{X,C}. \quad (A11)$$

Therefore

$$\vec{A}'^{X,C} = O\left(\vec{G}^{X,C}\right)^{-1} \cdot \vec{B}^{X,C} = O\vec{A}^{X,C}, \quad (A12)$$

and

$$\vec{A}'^{X,C} \cdot \vec{\Gamma}[n'](\mathbf{r}) = \vec{A}^{X,C}O^T \cdot O\vec{\Gamma}[n](M\mathbf{r}) = \\ \vec{A}^{X,C} \cdot \vec{\Gamma}[n](M\mathbf{r}). \quad (A13)$$

i.e. combining (A5) and (A13), $v_{X,C}^\sigma$ is fully GC.

For the TD case we restrict ourselves to TC only. Thus, consider two Cartesian systems, $\mathbf{r}'$ and $\mathbf{r}$, distinguished by a TD translational vector $\mathbf{x}(t)$, $\mathbf{r}' = \mathbf{r} + \mathbf{x}(t)$, the density and its CM are transformed as

$$n'_\sigma(\mathbf{r},t) = n_\sigma(\mathbf{r} + \mathbf{x}(t),t), \\ \mathbf{D}[n'] = \frac{1}{N_e(t)} \int d r n'(\mathbf{r},t)\mathbf{r} = \\ \frac{1}{N_e(t)} \int d r n(\mathbf{r} + \mathbf{x}(t),t)\mathbf{r} = \mathbf{D}[n] - \mathbf{x}(t). \quad (A14)$$

Thus, similarly to Eq. (A3),

$$\mathbf{l}_1^\sigma[n'](\mathbf{r},t) = \mathbf{l}_1^\sigma[n](\mathbf{r} + \mathbf{x}(t),t), \\ l_3^\sigma[n'](\mathbf{r},t) = l_1^\sigma[n](\mathbf{r} + \mathbf{x}(t),t). \quad (A15)$$

Hence, from the continuity equation,

$$\nabla \cdot \mathbf{j}'_\sigma(\mathbf{r},t) = \\ \nabla \cdot \mathbf{j}_\sigma(\mathbf{r} + \mathbf{x}(t),t) - \nabla n_\sigma(\mathbf{r} + \mathbf{x}(t),t) \cdot \dot{\mathbf{x}}(t), \quad (A16)$$

and together with the transformation of the CM velocity

$$\dot{\mathbf{D}}[n'] = \dot{\mathbf{D}}[n] - \dot{\mathbf{x}}(t), \qquad (A17)$$

we conclude that $l_2^\sigma$ transforms as a scalar:

$$\begin{aligned}
l_2'^\sigma(\mathbf{r},t) &= \dot{N}_\sigma(\mathbf{r},t) = -\nabla \cdot [\mathbf{j}'_\sigma(\mathbf{r},t) - n'_\sigma(\mathbf{r},t)\dot{\mathbf{D}}[n']] = \\
&-\{\nabla \cdot \mathbf{j}_\sigma(\mathbf{r}+\mathbf{x}(t),t) - \nabla n_\sigma(\mathbf{r}+\mathbf{x}(t),t) \cdot \dot{\mathbf{x}}(t)\} - \\
&-\{\nabla n_\sigma(\mathbf{r}+\mathbf{x}(t),t)\dot{\mathbf{D}}[n] - \nabla n_\sigma(\mathbf{r}+\mathbf{x}(t),t)\dot{\mathbf{x}}(t)\} = \\
&\dot{N}_\sigma(\mathbf{r}+\mathbf{x}(t),t) = l_2^\sigma(\mathbf{r}+\mathbf{x}(t),t).
\end{aligned} \qquad (A18)$$

Since the matrix $L(t)$ and the vector $\vec{B}^{X,C}$ (see Eq. (32)) are, again, constructed from spatial integration of TC terms, they transform as in the time-independent case, i.e.

$$\begin{aligned}
\vec{\vec{L}}'(t) &= \vec{\vec{L}}(t), \\
\vec{B}'(t) &= \vec{B}(t).
\end{aligned} \qquad (A19)$$

Therefore, repeating the same proof for the time-independent case, we can conclude that

$$\vec{A}'(t) \cdot \Lambda'_\sigma(\mathbf{r},t) = \vec{A}(t) \cdot \Lambda_\sigma(\mathbf{r}+\mathbf{x}(t),t), \qquad (A20)$$

and that TD $v_{XC}^\sigma$ in Eq. (30) is TC.

**APPENDIX B: AVOIDING SINGULARITY AND 'HYDRODYNAMICAL' INSTABILITY**

The singularity of Eqs. (32) arises from the matrix element $L_{d+1,d+1}(t) = \int dr \dot{N}(\mathbf{r},t)^2$, when $|\dot{N}(\mathbf{r},t)| \to 0$. Therefore, we regularize it by adding a small positive term, $\varepsilon(t)$, e.g.

$$\begin{aligned}
L_{d+1,d+1}^r(t) &= L_{d+1,d+1}(t) + \varepsilon(t), \\
\varepsilon(t) &= \max\{\varepsilon_0 / (1+t)^n, \varepsilon_{\min}\},
\end{aligned} \qquad (B1)$$

where the following values yielded good results: $\varepsilon_0 \sim 10^{-6} - 10^7$, $\varepsilon_{\min} \sim 10^{-8} - 10^{-10}$ and $n = 2$. Predicting the best regularization parameters is a nontrivial task. Therefore, in addition, we smoothly cut off the Lagrange multiplier $a_2(t) \equiv A_{d+1}(t)$, i.e.

$$a_2^c(t) = \begin{cases} a_2(t), & 0 \leq a_2(t) \leq \alpha_{\max}, \\ [1 + (a_2(t) - \alpha_{\max})/2a_2(t)]\alpha_{\max}, & a_2(t) > \alpha_{\max}, \\ 0, & a_2(t) < 0, \end{cases} \quad \text{(B2)}$$

We take $500 < \alpha_{\max} < 800$. It is also necessary to cut off $a_2(t)$ from below, to prevent negative values of $a_2(t)$. The reason is a 'hydrodynamical' instability which is related to the current-density equation of motion

$$d\mathbf{j}(\mathbf{r},t)/dt = T(\mathbf{r},t) - n(\mathbf{r},t)\nabla[v_{ext}(\mathbf{r},t) + v_H(\mathbf{r},t) + v_{XC}(\mathbf{r},t)],$$

where $T(\mathbf{r},t) = \langle i[\hat{T},\hat{\mathbf{j}}] \rangle$. In this equation the constrained $v_{XC}$ contributes a viscosity-like term: $a_2(t)n(\mathbf{r},t)\nabla[\nabla \cdot \mathbf{j}(\mathbf{r},t)]$, and thus it is clear that negative viscosity coefficient $a_2(t)$ values yield a numerical instability. We also confirm this behavior numerically, when $a_2(t)$ was not cut off from below. In this case, significant unphysical noise and amplification emerges in the very high harmonics of the emission spectra.

It should be obvious that following cutoffs, the EB condition will not be preserved exactly. Therefore, in order to compensate for this on the average, we replace $B_{d+1}(t)$ by $B_{d+1}^r(t) = (E(t) - W_{ext}(t))/\tau + B_{d+1}(t)$. Here $E(t)$ and $W_{ext}(t)$ are defined through Eq.(27), and $\tau \sim 0.1 - 1$ a.u. is recommended.

In summary, the calculation of the Lagrange multipliers is done as follows:

i) Solve Eqs. (32) following replacing $L_{d+1,d+1}(t)$ and $B_{d+1}(t)$ by $L_{d+1,d+1}^r(t)$ and $B_{d+1}^r(t)$, respectively.

ii) Calculate $a_2^c(t)$ from Eq. (B2), if $a_2(t)$ was cutoff by, recalculate the other Lagrange's multipliers for the fixed Lagrange's multiplier $A_{d+1}^c \equiv a_2^c$, from the linear equations

$$\tilde{\tilde{L}}(t)\vec{A}(t) = \tilde{\vec{B}}(t), \quad \text{(B3)}$$

where $\tilde{\tilde{L}}$ and $\overset{\leftrightarrow}{L}$ are identical except for the $d+1$ row; $\tilde{L}_{d+1,j}(t) = \delta_{j,d+1}$, and $\tilde{B}$ and $B$ are identical except for the $d+1$ entity; $\tilde{B}_{d+1}(t) = a_2^c(t)$.